# What is the Fastest Speed at Which a Single Electron Can Be Detected?

Deepak S. Rao[1], Thomas Szkopek[1], HongWen Jiang[2] and Eli Yablonovitch[1]

[1] UCLA Electrical Engineering Dept., Los Angeles CA 90095-1594

[2] UCLA Physics Dept., Los Angeles CA 90095

Abstract:

Electrometers measure electric charge, but there must be a fundamental speed limit to measuring one electric charge. Since there are no dimensional inputs to this question, the answer must be expressible in terms of the fundamental physical constants of Nature, e, $\hbar$, m, c. In general the question should be posed without reference to any specific technology, but for definiteness, we analyze the field effect transistor, which is essentially an electrometer. In spite of selecting a specific technology, we find that the speed limit is related to a fundamental constant, the Rydberg frequency, $me^4/2\hbar^3$, or as appropriate, the semiconductor Rydberg frequency $m^*e^4/2\hbar^3\varepsilon_r^2$, where $m^*$ is the electron effective mass, and $\varepsilon_r$ is the relative dielectric constant. We do not know whether the Rydberg frequency represents the upper speed limit, but on dimensional grounds we claim that the final limit can only differ by some function of the fine-structure-constant, $e^2/\hbar c$, or some other dimensionless fundamental constant.

**What is the Fastest Speed at which a Single Electron Can Be Detected?**

Millikan already showed[1] that it was possible to measure the charge on one electron. Nowadays we have much more sensitive instruments; the single electron transistor[2] (SET), the field effect transistor[3] (FET), and the quantum point contact[4,5] FET. It is not difficult to measure a single electronic charge.

There is ample motivation. In optical communications it is very desirable to be able to detect a single photo-electric charge[6]. Moreover, nano-electronics is rapidly approaching the one-electron limit. In spintronics, electron spin is measured by converting spin[7,8,9] to electric charge. The question arises, how quickly can we measure a change in electric charge by one unit?

The question is fundamental, since it is posed independently of technology, and is not constrained by any external parameters. Thus we should expect an upper limit that can be expressed in terms of fundamental constants, e, $\hbar$, m, c.

In this paper we treat the field effect transistor as an archetypal electrometer, and we ask: What is the fastest speed at which a single electric charge can be measured, with unity signal-to-noise ratio?

In spite of choosing specific technologies, we find that the speed/sensitivity limit is related to a fundamental constant, the Rydberg frequency $me^4/2\hbar^3$. This frequency corresponds to $\sim 2\times10^{-8}\,e/\sqrt{Hz}$ charge sensitivity, which is considerably faster and more sensitive than the best[2] that has thus far been demonstrated experimentally, in any technology.

In a semiconductor, the electric charge is screened by the relative dielectric constant $\varepsilon_r$, and the speed slows down according to the semiconductor Rydberg $m^*e^4/\varepsilon_r^2 2\hbar^3$ which corresponds to ~1 THz, and a sensitivity $\sim 10^{-6} e/\sqrt{Hz}$. This is more or less in line with the very best experimental results[2] in SET's, though SET's are not burdened by such a high dielectric screening. Thus further improvements in speed and sensitivity should be possible.

In this paper we analyze the following three cases: Case (1), an FET channel in the form of a cylindrical wire; Case (2), a quantum point contact in a pinched-off 2D electron gas; Case (3), an SET based on tunneling. We find that the speed limit tends to be roughly the Rydberg frequency in all three cases.

The field effect transistor is, at its simplest, a source/drain resistor whose conductance senses the presence of nearby electrostatic charge. Consider the example shown in Fig. 1(a). The thermal and shot noise are respectively; $I_N^2 = 4kT\langle G\rangle \Delta f + 2\langle I\rangle e\Delta f$, where $\langle I \rangle$ is the sensing current that monitors the conductance $\langle G \rangle$, and $\Delta f$ is the measurement bandwidth. It might be supposed that the speed/sensitivity limit might depend on temperature, but without loss of generality, the highest sensitivity is achieved when the thermal energy kT is negligible. Under the best possible circumstances, the amplitude signal-to-noise ratio (SNR) in the resistor is:

$$\text{SNR} = \frac{\langle I \rangle}{I_N} = \sqrt{\frac{I}{2e\,\Delta f}} = \sqrt{\frac{\langle G \rangle V_{DS}}{2e\,\Delta f}} \quad ،\dots\dots\dots\dots\dots\dots\dots(1),$$

that requires a sensing current $I = \langle G \rangle V_{DS}$ large enough to make the thermal noise negligible compared to the shot noise, which occurs when the bias voltage $V_{DS} > 2kT/e$.

A solid cylindrical semi-conducting wire, Case (1) as shown in Fig. 1(a), is an interesting channel geometry for estimating the ultimate charge sensitivity limit. The most favorable location, for the charge that needs to be detected, is in the center of the channel. It is perhaps easier to think of it as a negative ionic charge or a deep trap. The Coulomb potential of that ion, shown in Fig. 1(b), blocks the flow of current through the source/drain channel that is driven by the external source/drain voltage $V_{DS}$. To provide an upper detection speed limit, in the most optimistic circumstances, the presence/absence of this ion is assumed to fully switch off/on the channel current.

It might be supposed that increasing the channel radius R to enhance the conductance G would increase the signal to noise ratio, eq'n. (1), but then the ion coulomb potential $e/\varepsilon_r R$ at the periphery of the wire, Fig. 1(b), would be weaker and might become insufficient to pinch off the conductance. Likewise it might be supposed that the highest source/drain voltage $V_{DS}$ would be best, but then it could overwhelm the ion potential at the periphery of the wire, preventing pinch-off. *We show that the extraneous parameters, like channel radius R and bias voltage $V_{DS}$, tend to cancel out under optimal circumstances, leaving behind a sensitivity/speed limit that depends only upon fundamental physical constants.*

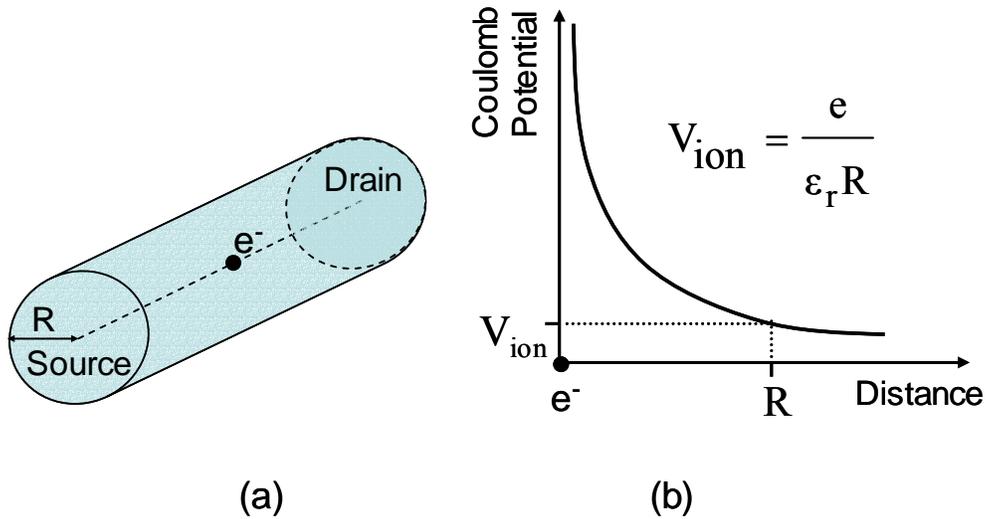

(a)　　　　　　　　　　(b)

Figure 1: (a) The FET channel in the form of wire with a 2D density of conducting modes. The charge to be detected, is an immobile ion/electron placed at the center of the channel. (b) Ionic Coulomb potential roll-off along the radius of the conductor.

Based on the above discussion, $V_{DS}$ should equal no more than $e/\varepsilon_r R$. Let us now estimate the highest possible channel conductance G that can be used. G is at its maximum when the channel transport is ballistic, and the only effective scattering center is the fixed ionic charge that pinches off the channel. In the ballistic regime, the average channel conductance $\langle G \rangle$ in the conducting state is due to $N_m$ conducting modes in the channel, each contributing a conductance of $e^2/h$ to channel. The conductance $e^2/h = 1/26$ k$\Omega$ is the well-known standard quantum of conductance of an electron waveguide mode in a 1-dimensional channel. It is presumed that the wire transports freely in the longitudinal direction and is limited to a finite number $N_m$ of conducting modes in the transverse directions. The well-known transverse 2-dimensional density of states $m^*/\pi\hbar^2$ per unit area, per unit energy, including 2 spin degenerate states, becomes:

$$N_m = \left(\frac{m^*}{\pi\hbar^2}\right)(\pi R^2)E_{kin} = \left(\frac{m^*}{\pi\hbar^2}\right)(\pi R^2)eV_{DS} = \left(\frac{m^*}{\hbar^2}\right)\left(\frac{e^2}{\varepsilon_r}\right)R \quad \ldots\ldots\ldots\ldots(2),$$

where the available kinetic energy $E_{kin}$ of the electrons multiplies the density of states per unit energy. Under ballistic conditions at low temperatures, $E_{kin} = eV_{DS} = e^2/\varepsilon_r R$ at the maximum permitted source/drain voltage $V_{DS}$. Then $N_m$ becomes equal to $(m^*/\hbar^2)(e^2/\varepsilon_r)R$ on the right side of eq'n. (2). This is multiplied by the conductance quantum $e^2/h$ to get conductance, and by $V_{DS}$ to get the total sensing current I:

$$I = \left(\frac{e^2}{h}\right)\left(\frac{m^*}{\hbar^2}\right)\left(\frac{e^2}{\varepsilon_r}\right)R\left(\frac{e}{\varepsilon_r R}\right) = \left(\frac{e}{h}\right)\left(\frac{m^* e^4}{\hbar^2 \varepsilon_r^2}\right) = 2e \times \frac{\text{Rydberg}}{h}\left(\frac{m^*}{m\varepsilon_r^2}\right) \quad \ldots\ldots\ldots\ldots(3)$$

where the Rydberg is the ordinary hydrogenic Rydberg binding energy or the semiconductor Rydberg energy that is scaled by the factor $m^*/m\varepsilon_r^2$ as dictated by the dielectric screening $\varepsilon_r$, and by effective mass ratio $m^*/m$. We see that the current in eq'n. (3) is simply the electric charge multiplied by appropriate Rydberg frequency. Thus the signal-to-noise ratio becomes:

$$\text{SNR} = \frac{\langle I \rangle}{I_N} = \sqrt{\frac{2e \times \text{Rydberg}}{2eh\,\Delta f}} = \sqrt{\frac{\text{Rydberg}}{h\,\Delta f}} \quad or \quad \sqrt{\frac{\text{Rydberg}}{h\,\Delta f}\left(\frac{m^*}{m\varepsilon_r^2}\right)} \quad \ldots\ldots\ldots\ldots (4).$$

The maximum bandwidth for unity signal-to-noise ratio detection of a single charge is the Rydberg frequency. It should be noted that the Rydberg energy is composed of fundamental constants: $(\frac{1}{2})mc^2 \times (e^2/\hbar c)^2$, which is half the electron rest energy times the fine structure constant squared.

We now consider Case (2), a more traditional FET geometry, namely the quantum point contact transistor[4] that has become a very popular tool in mesoscopic physics experiments. A quantum point contact (QPC) is a ballistic constriction in a transistor channel, with a width comparable to the electron Fermi wavelength. A convenient way of creating a QPC is by electrostatically confining a 2D electron gas in a modulation doped heterostructure between split gate electrodes as shown in Fig. 2(a). The key feature in the transport through a QPC is the quantization of conduction in multiples of $e^2/h$ due the formation of 1D waveguide modes in the narrow channel. For this Case (2), we find a similar universal signal-to-noise ratio as eq'n. (4), except as modified by geometrical factors.

We consider the sensitivity of a QPC with a width W equal to just one-half of a de Broglie wavelength so that there is only one electron waveguide mode through the constriction. As in the previous Case (1), the channel is assumed to be completely pinched off in case of the presence of a single ionic charge in the transport path. The waveguide mode is opened when the charge is absent, turning the conductance on by one full quantum $e^2/h$. To distinguish Case (2) from the previous Case (1), the maximum permitted applied source-drain voltage is taken to be equal to the 1→2 sub-band spacing, $\Delta_{\text{sub-band}}$. Accounting for spin-degeneracy, the signal-to-noise ratio expression in eq'n. (1) becomes

$$\text{SNR} = \frac{\langle I \rangle}{I_N} = \sqrt{\frac{2e^2 V_{DS}}{2h \, e \, \Delta f}} = \sqrt{\frac{\Delta_{\text{sub-band}}}{h \, \Delta f}} \quad \ldots\ldots\ldots\ldots\ldots\ldots\ldots\ldots(5)$$

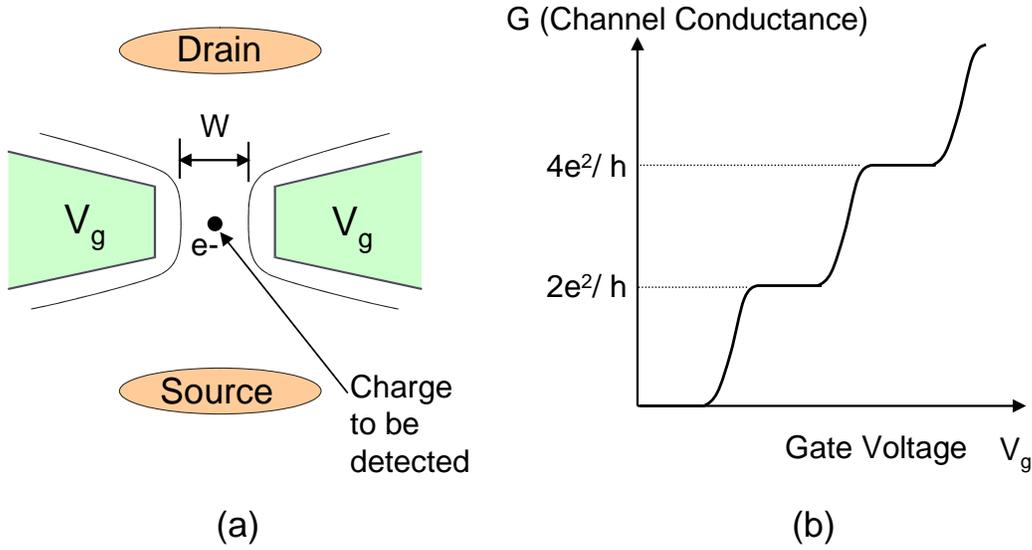

(a)                                (b)

Figure 2: (a) A quantum point contact field effect transistor patterned by electrostatic gates on a 2d electron gas. W is the width of the conducting channel. The charge to be detected is optimally located at the center of the transistor. (b) The effect of the charge is to shift the channel conductance by $2e^2/h$, the conductance quantum, allowing 2 spin states.

An estimate for $\Delta_{\text{sub-band}}$ can be made by using the 1→2 energy level spacing for a waveguide of width W, which gives $\Delta_{\text{sub-band}} = 3\pi^2 \hbar^2 / 2m^* W^2$. Substituting this value of $\Delta$ in the SNR expression of eq'n. (5)

$$\text{SNR} = \frac{\langle I \rangle}{I_N} = \sqrt{\frac{\Delta_{\text{sub-band}}}{h \, \Delta f}} = \sqrt{\frac{3\pi^2 \hbar^2}{2m^* W^2} \frac{1}{h \, \Delta f}} \quad \ldots\ldots\ldots\ldots\ldots(6)$$

Eq'n. (6) can be re-expressed in terms of the vacuum Bohr radius $a_o \equiv \hbar^2/me^2$ and the vacuum Rydberg:

$$\text{SNR} = \sqrt{\frac{3\pi^2 \hbar^2}{2m^* h \Delta f}\left(\frac{me^2}{\hbar^2}\right)^2 \frac{a_o^2}{W^2}} = \sqrt{\frac{3\pi^2}{h\Delta f}\left(\frac{m}{m^*}\right)\frac{me^4}{2\hbar^2}\frac{a_o^2}{W^2}} = \sqrt{\frac{3\pi^2 \text{Rydberg}}{h\Delta f}\left(\frac{m}{m^*}\right)\left(\frac{a_o}{W}\right)^2} \quad\ldots\ldots(7)$$

The SNR for a quantum point contact transistor, eq'n. (7), resembles the SNR for a cylindrical transistor channel, eq'n. (4), but with some differences, especially that the dielectric screening $(1/\varepsilon_r)^2$ is replaced by $(a_o/W)^2$.

Let us now analyze Case (3), single electric charge detection in an SET. In Fig. 3, a conducting island is linked by tunnel barriers to a source, and a drain electrode. Due to the Coulomb blockade[10], no current will flow unless the source/drain voltage is sufficient to overcome the single electron charging energy, $e^2/2C$, of the island capacitance C. Since an external electrode can adjust the exact electric potential of the island, the external gate can switch[2] the blockade effect on and off, making a transistor, and an electrometer.

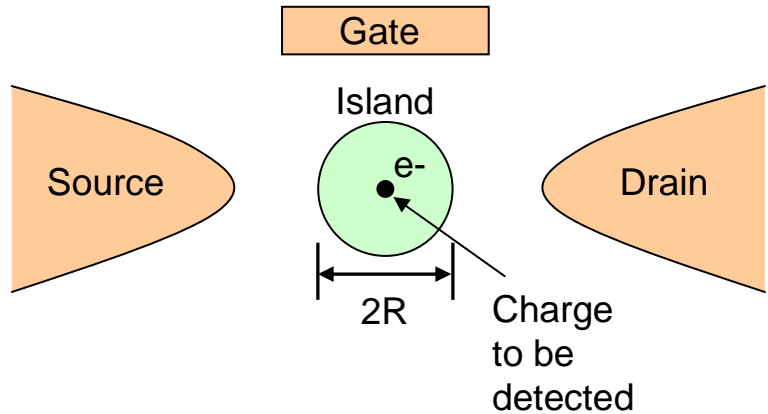

Figure 3: A Single Electron Transistor formed by an isolated metallic Island between Source and Drain reservoirs. R is the radius of the Island. The charge to be detected is optimally located at the center of the Island.

Just as in eq'n. (1), it is desirable to make the sensing current $\langle G \rangle V_{DS}$ as large as possible. However, the tunneling conductance $\langle G \rangle$, should not exceed the conductance quantum $e^2/h$, or the charging energy would become ill-defined. Furthermore the source/drain voltage $V_{DS}$ should not exceed the Coulomb blockade voltage $e/2C$. Thus SNR is limited by:

$$SNR = \frac{\langle I \rangle}{I_N} = \sqrt{\frac{\langle G \rangle V_{DS}}{2e\Delta f}} = \sqrt{\frac{2e^2}{2eh\Delta f}\frac{e}{2C}} = \sqrt{\frac{e^2}{h\Delta f}\frac{1}{2C}} \quad \ldots\ldots\ldots(8)$$

where eq'n. (8) is actually similar to the corresponding SET charge sensitivity equation in ref. 2, except for geometrical factors. The capacitance C can be estimated[11] by regarding the island as a thin conducting disk, whose capacitance is $C=2\varepsilon_r R/\pi$. Then the SNR becomes:

$$SNR = \frac{\langle I \rangle}{I_N} = \sqrt{\frac{e^2}{2h\Delta f}\frac{\pi}{2\varepsilon_r R}} = \sqrt{\frac{e^2}{h\Delta f}\frac{me^2}{2\hbar^2}\frac{\pi a_o}{2\varepsilon_r R}} = \sqrt{\frac{Rydberg}{h\Delta f}\frac{\pi a_o}{2\varepsilon_r R}} \quad \ldots\ldots\ldots(9)$$

The charge detection speed at SNR=1, is again the Rydberg frequency in eq'n. (9), but eq'n. (9) now incorporates $a_o/\varepsilon_r R$, the geometric mean of the dimensionless charge screening correction factors from eq'n. (4) and eq'n. (7), provided R and W have the same lithographic size limit. In practice, the numerical value of the approximate correction factors $(1/\varepsilon_r)^2$, $(a_o/W)^2$, and $(a_o/\varepsilon_r R)$ from equations (4), (7), and (9) respectively, are all very similar. Thus the speed with which a single electric charge can be measured is essentially limited by a fundamental constant, the Rydberg frequency, as corrected by charge screening or geometrical factors. The sensitivity/speed limit can be $10^{-7}$ -$10^{-6}$ e/√Hz depending on charge screening and geometry.

Thus Case (1), the FET, Case (2), the quantum point contact FET, and Case (3), the SET, all have similar speed/sensitivity limits connected to the Rydberg frequency. The question arises whether a different technology, using perhaps high energy particles could have a better speed limit. We believe that it may be possible to measure more quickly, but on dimensional grounds the improved speed would still scale as the Rydberg frequency, as modified by a dimensionless factor, like some power of the fine structure constant.

The work is supported by the Defense Advanced Research Projects Agency (MDA972-99-1-0017), Army Research Office (DAAD19-00-1-0172) & the Defense Micro-Electronics Activity.